\documentclass{PoS}
\usepackage{url}

\title{Interpreting LHC searches for new physics with SModelS}

\ShortTitle{Interpreting LHC searches for new physics with SModelS}

\author{\speaker{Ursula Laa}%
       \thanks{In collaboration with Chiara Arina, Genevi\`eve B\'elanger, Maria Eugenia Cabrera Catalan, Jonathan Da Silva, Sabine Kraml, Suchita Kulkarni, Andre Lessa, Veronika Magerl, Wolfgang Magerl, Doris Proschofsky-Spindler, Alexander Pukhov, Michael Traub and Wolfgang Waltenberger.} \\
       Laboratoire de Physique Subatomique et de Cosmologie, Universit\'e Grenoble-Alpes, CNRS/IN2P3, 53 Avenue des Martyrs, F-38026 Grenoble, France\\
LAPTH, Universit\'e Savoie Mont Blanc, CNRS, B.P.110, F-74941 Annecy-le-Vieux Cedex, France\\
       E-mail: \email{ursula.laa@lpsc.in2p3.fr}}

\abstract{ATLAS and CMS have performed a large number of searches for physics beyond the Standard Model (BSM). The results are typically presented in the context of Simplified Model Spectra (SMS), containing only a few new particles with fixed decay branching ratios, yielding generic upper limits on the cross section as a function of particle masses. The interpretation of these limits within realistic BSM scenarios is non-trivial and best done by automated computational tools.
To this end we have developed SModelS, a public tool that can test any given BSM model with a $\mathbb{Z}_2$ symmetry by decomposing it into its SMS components and confronting them with a large database of SMS results. This allows to easily evaluate the main LHC constraints on the model. Additionally, SModelS returns information on important signatures that are not covered by the existing SMS results. This may be used to improve the coverage of BSM searches and SMS interpretations.
We present the working principle of SModelS, in particular the decomposition procedure, the database and matching of applicable experimental results. Moreover, we present applications of SModelS to different models: the MSSM, a model with a sneutrino as the lightest supersymmetric particle and the UMSSM. These results illustrate how SModelS can be used to identify important constraints, untested regions and interesting new signatures.}

\FullConference{The European Physical Society Conference on High Energy Physics\\
         22-29 July 2015\\
         Vienna, Austria}

\begin{document}

\section{Introduction}

Simplified Model Spectra are an effective Lagrangian description, containing a limited set of new particles with fixed decay modes.
The free parameters of Simplified Models are just the masses of the new particles, the decay branching ratios typically being set to $100$\%.
A large number of Run 1 LHC searches for new physics have been interpreted in terms of SMS,
giving upper limits on the production cross section as a function of the masses of the new particles in the SMS (typically in 2-dimensional mass planes).
The interpretation of these limits in terms of complete BSM models is, however, not trivial.
SModelS~\cite{Kraml:2013mwa,Kraml:2014sna} is an automated tool that decomposes a realistic model into SMS components and matches them to corresponding results in the database.
The decomposition procedure is generic and can be used for any BSM model presenting a $\mathbb{Z}_2$ symmetry.
Using this approach, one can test theory predictions against experimental results without running a Monte Carlo simulation.
Another advantage is that SModelS comes with a large internal database and it is easy to add new results.
The public database of SModelS v1.0 includes 21 ATLAS and 41 CMS SUSY search results at $8$~TeV.
In addition, SModelS can identify signatures which are not covered by the SMS results in the database, labeled as ``missing topologies''.
SModelS is a public tool, the program is available for download at \url{http://smodels.hephy.at}.

\section{Working Principle of SModelS}
\label{sec:wp}
\begin{figure}[h!]\centering
     \includegraphics[width=0.8\textwidth]{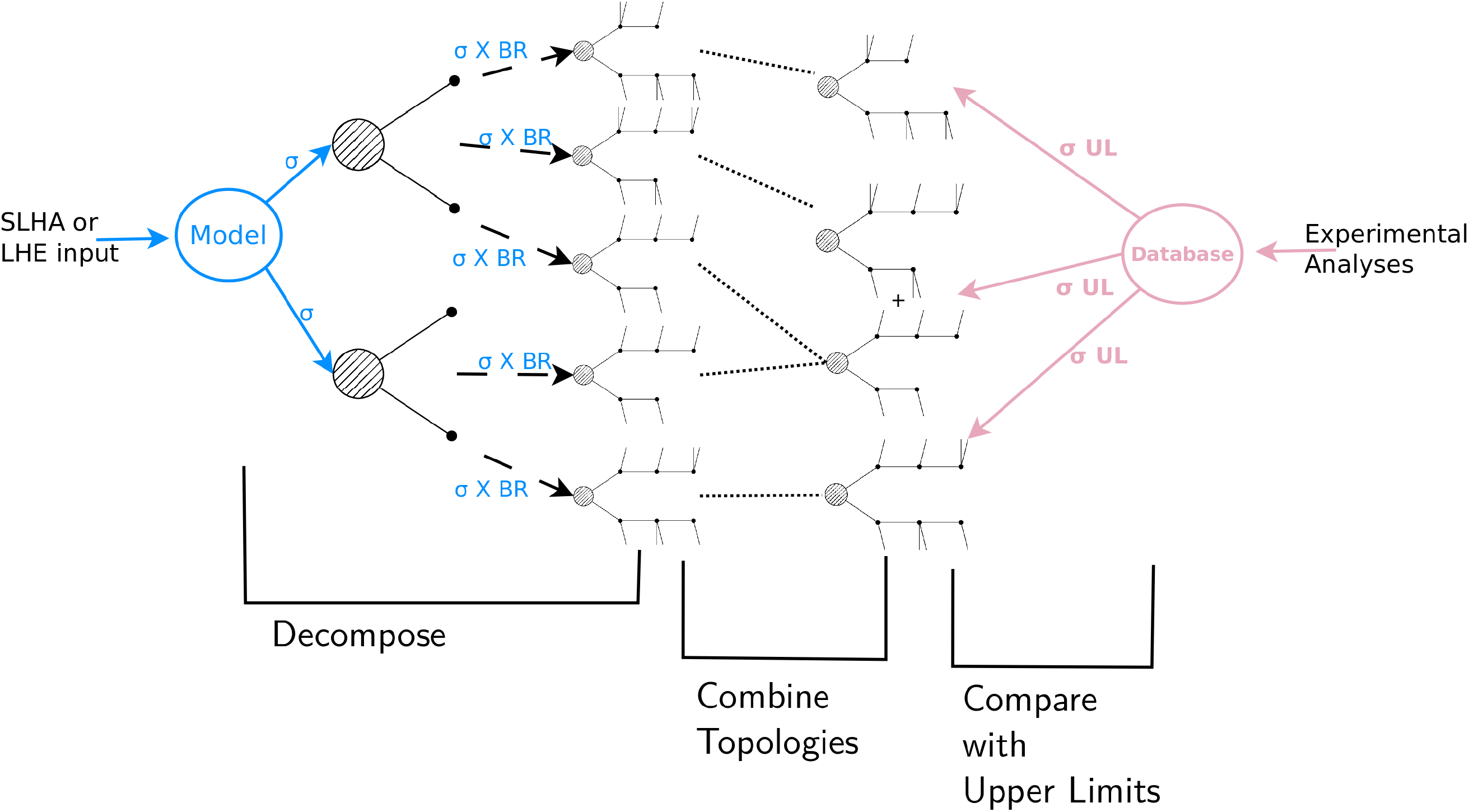}
     \caption{SModelS working principle\label{fig:workingPrinciple}}
\end{figure}

We assume that, in first approximation, SMS results depend only on the mass spectrum of the new particles, and not on specifics of the model (such as the production process or the spin structure).
In this case, topologies are fully described by the outgoing Standard Model (SM) particles in each vertex and the masses of the BSM states.
Any additional information on the new particles is discarded.
Topologies can then be described in the SModelS bracket notation.
The structure is \texttt{[branch1, branch2]} for the decay chains (``branches'') of the initially produced particles.
Each branch contains inner brackets for each vertex, containing in turn the list of outgoing SM particles.
An example is stop pair production, that can give rise to various SMS topologies.
A scenario where each stop decays directly, $\tilde{t}\rightarrow t \tilde{\chi}^0_1$, is described as  \texttt{[[[t]],[[t]]]}.
Alternatively, if each stop decays via a chargino, $\tilde{t}\rightarrow b \tilde{\chi}^{\pm}, \tilde{\chi}^{\pm}\rightarrow W \tilde{\chi}^0_1$, the topology is described as \texttt{[[[b],[W]],[[b],[W]]]}.
\\
Given this abstract notation of topologies, an input model specified e.g. in an SLHA file can be decomposed into SMS topologies.
For each topology SModelS keeps track of the masses of the new particles involved and calculates a weight $\sigma\times$BR.
All SMS components contributing to the same experimental result are combined, and
the corresponding sum of weights can then be compared directly against the upper limit
for the given mass combination.
This procedure is illustrated in Fig.~\ref{fig:workingPrinciple}.
A more detailed explanation can be found in \cite{Kraml:2013mwa}.
A topology which does not match any of the experimental results in the database is considered a missing topology.

\section{Application to the MSSM}

\begin{figure}[h!]\centering
\includegraphics[width=0.42\textwidth]{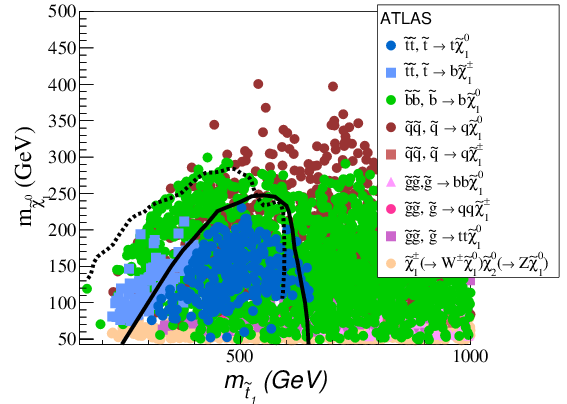}
\includegraphics[width=0.42\textwidth]{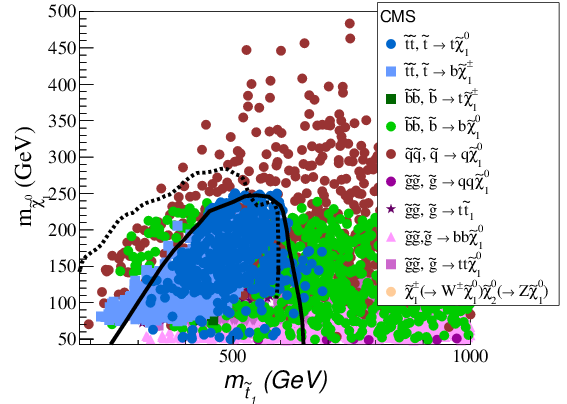}
\includegraphics[width=0.42\textwidth]{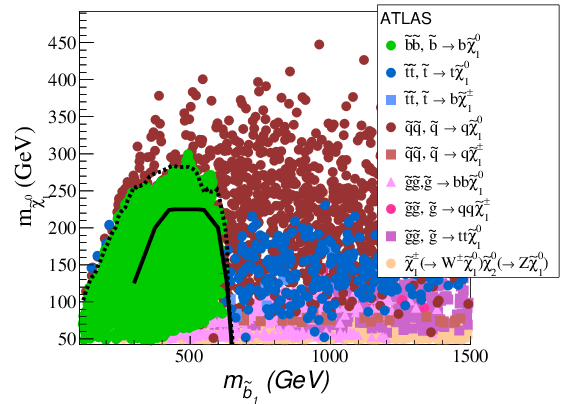}
\includegraphics[width=0.42\textwidth]{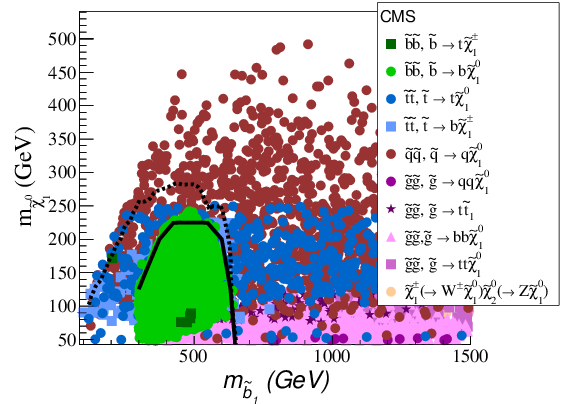}
\caption{The most constraining analysis is indicated for all excluded points of Scan-II. On the left for ATLAS, on the right for CMS results, in the mass planes of $\tilde{\chi}_1^{0}$ vs $\tilde{t}$ (top row) and $\tilde{b}$ (bottom row).\label{fig:scan2}}
\end{figure}

SModelS was applied in~\cite{Kraml:2013mwa} in two scans over generic MSSM parameter spaces, with parameters defined at the weak scale.
In both scans, an approximate GUT relation was employed for the gaugino masses, $M_1:M_2:M_3 = 1: 2: 6$.
The other free parameters are $\mu$, $\tan\beta$, and the squark and slepton mass parameters and trilinear couplings.
The parameter space of each respective scan can be summarized as
\begin{itemize}
\item Scan-I: light gauginos and sleptons, all squarks are heavy
\item Scan-II: light squarks (in particular stop, sbottom), heavy sleptons.
\end{itemize}
To illustrate the results we first show in Fig.~\ref{fig:scan2} how important constraints for specific regions of parameter space can be identified.
For each excluded point of Scan-II the most constraining analysis (by ATLAS or CMS respectively) is plotted in the $\tilde{\chi}_1^{0}$ vs $\tilde{t}$ or $\tilde{b}$ mass plane.
Additionally, exclusion lines from corresponding experimental SMS results are overlayed.
We observe good agreement between the SMS exclusion and the excluded $\tilde{t}$ and $\tilde{b}$ masses in the full model.

\begin{figure}[h!]\centering
     \includegraphics[width=0.48\textwidth]{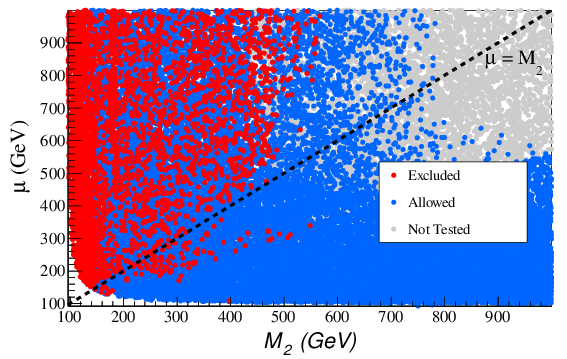}
     \includegraphics[width=0.48\textwidth]{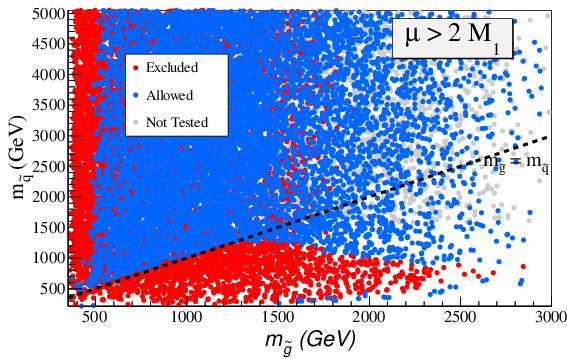}
\caption{Breakdown of the parameter points into excluded, allowed, and not tested points.
Points of Scan-I are shown in the plane of $M_2$ vs $\mu$ (left), points of Scan-II in the mass plane of $\tilde{g}$ vs $\tilde{q}$ (right).\label{fig:weak}}
 \end{figure}

SModelS can further be used to test to which extend certain regions of parameter space are constrained by SMS results, thus identifying regions
that are fully excluded, excluded only for certain parameter combinations or unconstrained by SMS results.
This is illustrated in Fig.~\ref{fig:weak} by grouping the scan parameter points that are excluded, allowed or not tested by SModelS.
On the left, points of Scan-I are shown in the plane of $M_2$ vs $\mu$.
Excluded points are found mainly for wino-like $\tilde{\chi}_1^{\pm}$, i.e. when $M_2 < \mu$.
On the right, we show points of Scan-II in the mass plane of $\tilde{g}$ vs $\tilde{q}$.
Points with heavy gluinos ($m_{\tilde{g}} > 500$ GeV) often evade constraints if $m_{\tilde{g}} < m_{\tilde{q}}$.
The reason is that in this region the gluino decays can be a mix of
$\tilde g\to q\bar q\chi^0_{1...4}$, $\tilde g\to q\bar
q'\chi^\pm_{1,2}$, or even $\tilde g\to g\chi^0_{1...4}$, a situation
that is poorly constrained by the current SMS results.

\section{Application to a Sneutrino Dark Matter Model}

\begin{figure}[t!]\centering
     \includegraphics[width=0.52\textwidth]{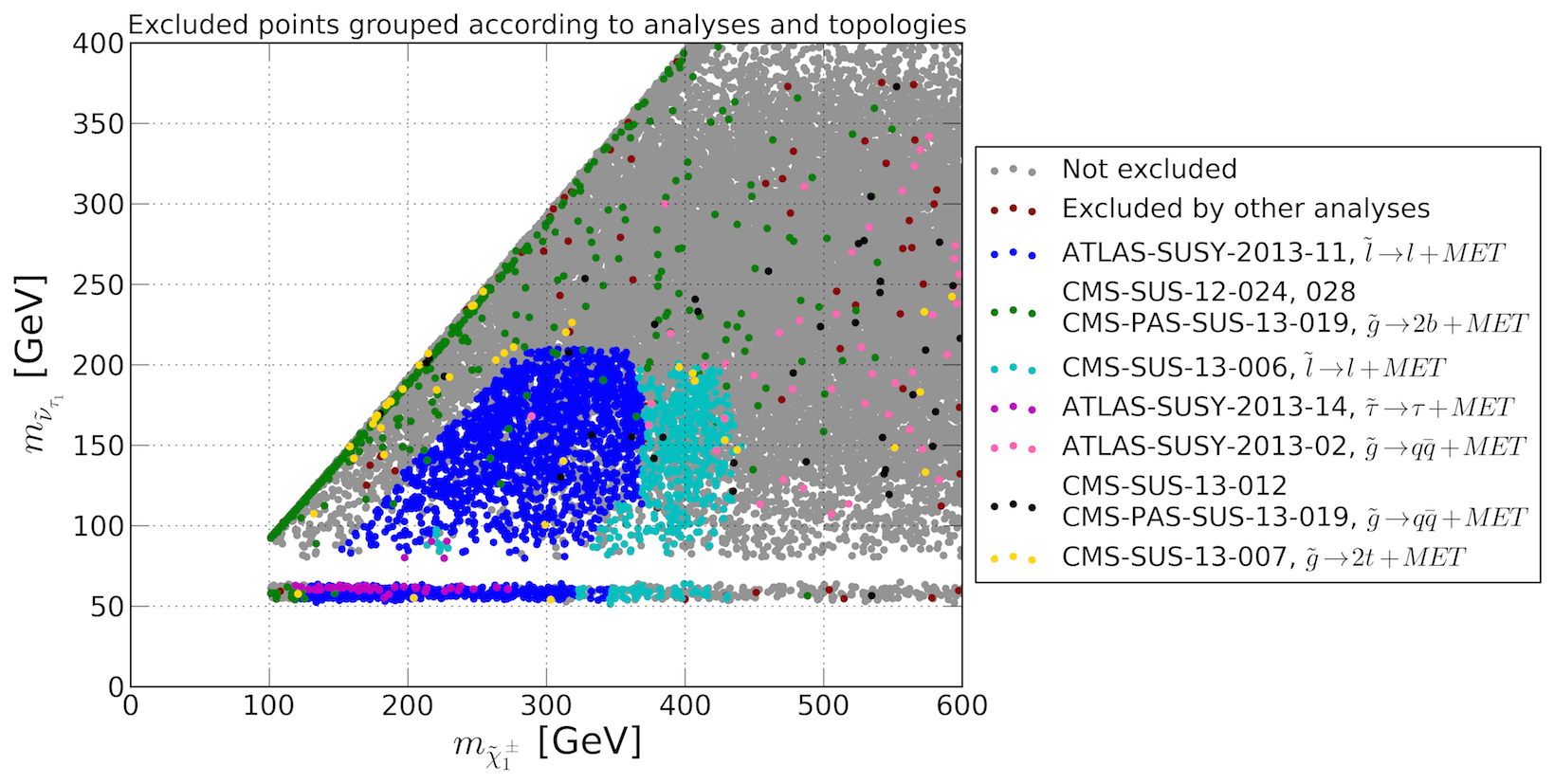}
\includegraphics[width=0.47\textwidth]{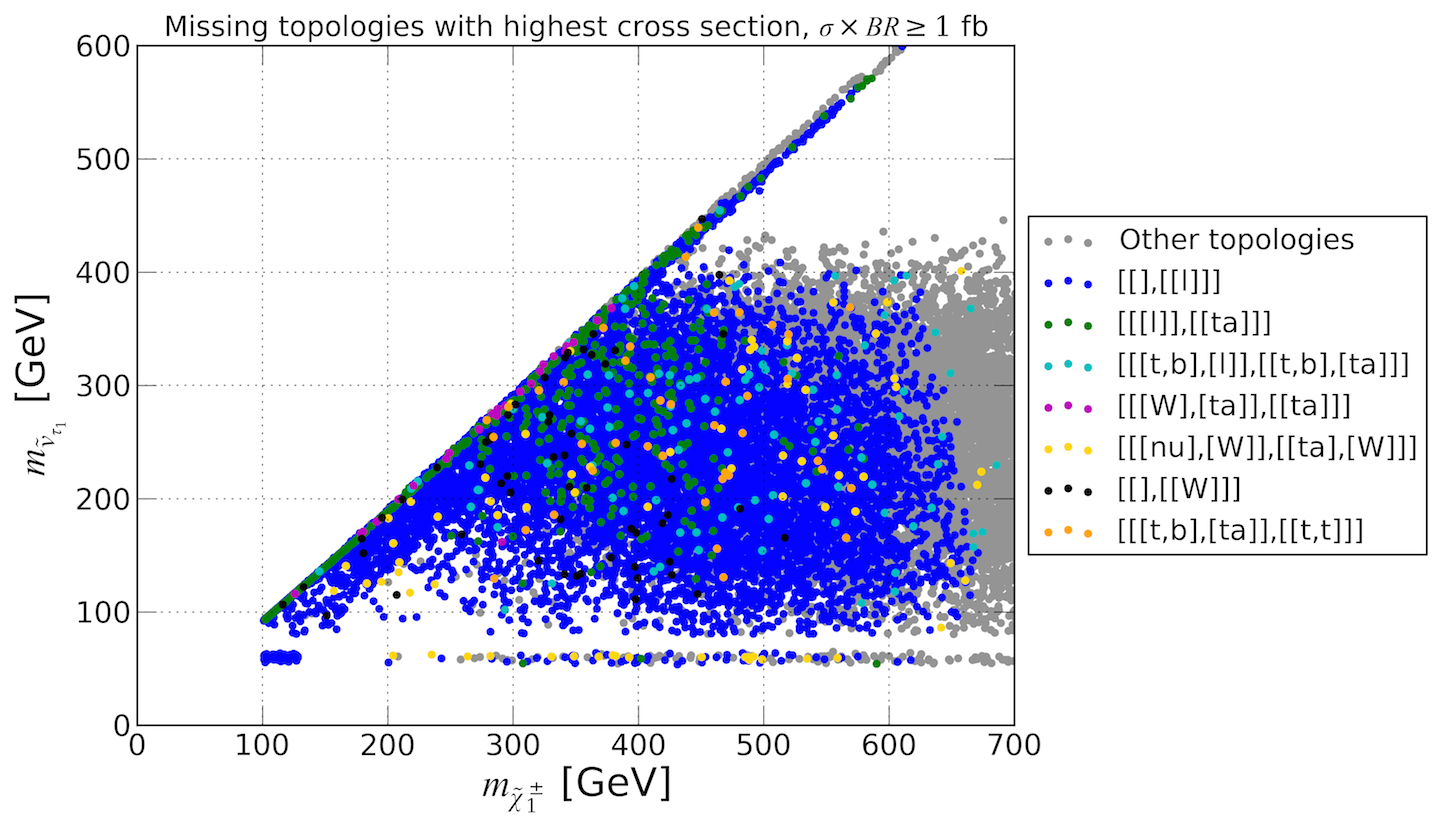}
\caption{In the mass plane of $\tilde{\chi}_1^{\pm}$ vs $\tilde{\nu}_{\tau_1}$ we show the most constraining analysis for excluded scan points (left) and for each unexcluded point the missing topology with the highest cross section (right).\label{fig:cha}}
\end{figure}

As argued above, the application of SModelS is not limited to the MSSM, and we can thus study
constraints from LHC Run 1 on models with different particle content.
In particular the lightest supersymmetric particle (LSP) need not be the $\tilde{\chi}_1^{0}$.
In~\cite{Arina:2015uea}, we studied the LHC Run 1 constraints for a sneutrino dark matter model, concretely the model of~\cite{Borzumati:2000mc}, where a right-handed (RH) neutrino superfield is added to the MSSM with a weak scale
trilinear coupling that is not proportional to the neutrino Yukawa coupling.
The RH sneutrino field can mix with the LH partner, and a mixed mainly RH sneutrino is then a viable dark matter candidate.
Such a scenario can lead to LHC signatures which are quite distinct from those of a neutralino LSP~\cite{Arina:2015uea}.
\\
Here we are discussing the results in the mass plane of $\tilde{\chi}_1^{\pm}$ and $\tilde{\nu}_{\tau_1}$, shown in Fig.~\ref{fig:cha}.
On the left we show, for each excluded point, the most constraining analysis.
Results from dilepton searches (obtained in the context of slepton pair production, followed by $\tilde{l}\rightarrow l \tilde{\chi}_1^{0}$) constrain chargino-pair production, where $\tilde{\chi}_1^{\pm}\rightarrow l^{\pm} \tilde{\nu}_{l_1}$ and exclude points with light $\tilde{\chi}_1^{\pm}$ and $\tilde{\nu}_{\tau_1}$.
On the right we show, for unexcluded points, the missing topologies with the highest cross section. 
The dominant missing topology is a single lepton signature deriving from chargino-neutralino production, where $\tilde{\chi}_1^{\pm}\rightarrow l \tilde{\nu}_{l_1}$ as seen before and $\tilde{\chi}_1^{0}\rightarrow \nu \tilde{\nu}$ is fully invisible,
\texttt{[[],[[l]]]}.

\section{Application to U(1) extensions of the MSSM (UMSSM)}

\begin{figure}[h!]\centering
     \includegraphics[width=0.447\textwidth]{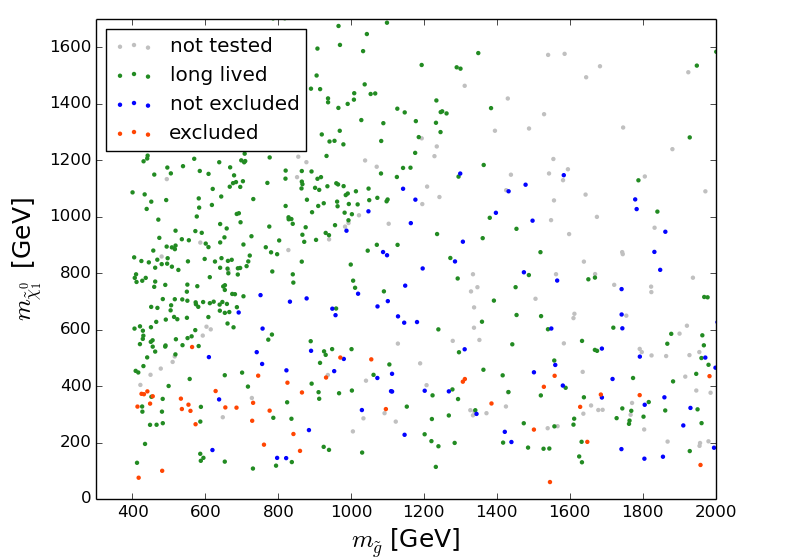}
     \includegraphics[width=0.545\textwidth]{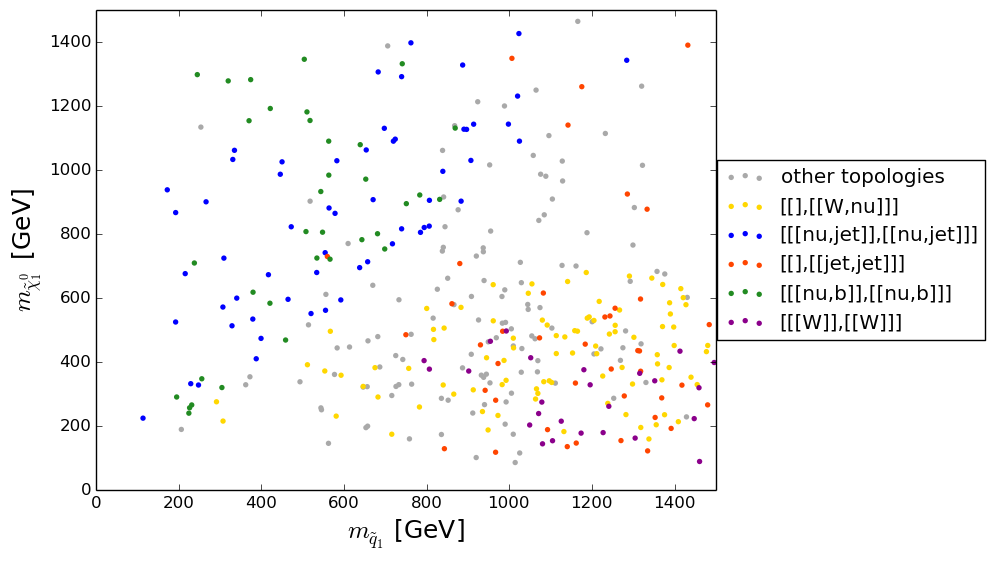}
\caption{A breakdown of scan points with a sneutrino LSP into not tested, allowed and excluded points and points with long-lived particles is shown in the mass plane of $\tilde{g}$ vs $\tilde{\chi}_1^{0}$ (left). For each unexcluded point the missing topology with the highest cross section is shown in the mass plane of $\tilde{q}$ vs $\tilde{\chi}_1^{0}$ (right).\label{fig:u1}}
\end{figure}

We further consider SMS constraints on a U(1) extended model, where the RH neutrino superfield is charged under the new gauge group~\cite{Belanger:2015cra}.
In this model, a purely RH sneutrino can be a thermal dark matter candidate.
Here we are focussing only on scan points with such a RH sneutrino LSP.
In this case, the collider signatures will depend strongly on the nature of the next-to LSP (NLSP).
Additional care must be taken because at the time this study was performed available SMS constraints applied only for prompt decays.
Parameter points that feature long-lived particles that would lead to distinct detector signals are thus flagged and not tested by SModelS.
This is illustrated in Fig.~\ref{fig:u1} (left), where we show scan points in the $\tilde{g}$ vs $\tilde{\chi}_1^{0}$ mass plane.
Points with long-lived charged particles, not tested by SModelS, are shown in green.
If the $\tilde{g}$ is lighter than the $\tilde{\chi}_1^{0}$ it is likely long-lived on detector scales.
Figure~\ref{fig:u1} (right) specifies, for unexcluded points in the mass plane of $\tilde{q}$ vs $\tilde{\chi}_1^{0}$, the missing topology with the highest cross section.
We outline two special topologies for which the signatures differ from those expected in a neutralino LSP scenario.
First, if the $\tilde{q}$ is lighter than $\tilde{\chi}_1^{0}$, additional neutrinos are found in the vertex,
$\tilde{q} \rightarrow q \nu_R \tilde{\nu}_R$.
This topology is denoted as 
 \texttt{[[[nu,jet]],[[nu,jet]]]} or \texttt{[[[nu,b]],[[nu,b]]]} for light quarks and b quarks respectively.
This resembles the dijet final state found in the MSSM.
However, the additional neutrinos will alter the event kinematics and can thus lead to different efficiencies.
Therefore upper limits obtained in the context of the MSSM might not apply.
Second, if the $\tilde{\chi}_1^{0}$ is light, $\tilde{\chi}_1^{\pm} \tilde{\chi}_1^{0}$ production can yield large cross sections for a single $W$ signature, \texttt{[[],[[W,nu]]]}.
This topology arises when both chargino and neutralino decay directly to the LSP, $\tilde{\chi}_1^{\pm}\rightarrow W \nu_R \tilde{\nu}_R$ and $\tilde{\chi}_1^{0}\rightarrow \nu_R \tilde{\nu}_R$.

\section{Conclusion and Outlook}
SModelS is a useful tool to test whether a particular scenario is excluded by the SMS results from ATLAS and CMS.
Apart from classifying the most important constraints, SModelS can also be used to identify untested regions, and find missing topologies as well as new signatures that might be interesting to look for.
Here we have presented results obtained with SModelS for three different models.
SModelS was also applied to other scenarios in, e.g.,~\cite{Barducci:2015zna}.
Future developments foresee for example the use of efficiency maps (work in progress).
An extension of SModelS treating signatures of heavy stable charged particles was recently presented in~\cite{Heisig:2015yla}.

\acknowledgments
UL acknowledges support by the ``Investissement d'avenir, Labex ENIGMASS'', the Theory-LHC-France initiative of CNRS (INP/IN2P3) and the ANR-12-BS05-0006 project DMAstroLHC.

\end{document}